\documentstyle[12pt,epsf,epsfig]{article}
\newcommand{\nn}{\nonumber}
\newcommand{\be}{\begin{equation}}
\newcommand{\ee}{\end{equation}}

\def\vk{{\bf k}_{\perp}}

\def\vb0{{\bf b}_0}

\def\als{\alpha_s}

\def\gev{\,{\rm GeV}}

\def\xb{\overline{x}}

\def\veps{\varepsilon}
\begin{document}

\begin{center}
{\bfseries ELECTROPRODUCTION OF VECTOR MESONS AT SMALL $x$}

\vskip 5mm

S.V.\ Goloskokov$^{\dag}$

\vskip 5mm

{\small {\it BLTP, Joint Institute for Nuclear Research \\
Dubna 141980, Moscow region, Russia}
\\
$\dag$ {\it E-mail:goloskkv@thsun1.jinr.ru }}
\end{center}

\vskip 5mm

\begin{center}
\begin{minipage}{150mm}
\centerline{\bf Abstract} Vector meson electroproduction is
analyzed within the two-gluon (2G) model and the generalized
parton distribution (GPD) approach at small $x$-Bjorken. We
demonstrate that 2G and GPD models are not completely equivalent.
At the same time, both models  are in reasonable agreement with
available experimental data on light vector meson
electroproduction.
\end{minipage}
\end{center}

\vskip 10mm
\section{Vector meson production in 2G and GPD models}
 This report is devoted to investigation  of vector meson
electroproduction at small Bjorken $x$ and large photon
virtuality. In the low - $x$ region  the predominant contribution
to the process is determined by the 2G exchange and the vector
meson is produced via the photon-two-gluon fusion.  At large $Q^2$
the cross section for the vector meson production is dominated by
the $\gamma _{L}^{\ast }\rightarrow V_{L}$ amplitude which
factorizes \cite{Ji97} into a  hard meson photoproduction off
gluons, and GPD. The amplitudes of $\gamma _{\perp }^{\ast
}\rightarrow V_{\perp }$ and the $\gamma _{\perp }^{\ast
}\rightarrow V_{L}$ transitions which are important in polarized
observables are suppressed as a power of $1/Q$ and exhibit
infrared singularities \cite{mp}. Similar properties of vector
meson production amplitudes were found within the 2G model by
several authors \cite{2gm}. Calculation of these higher twist
amplitudes requires a regularization scheme which depends on a
model. The modified perturbative approach (MPA) \cite{sterman}
which includes the transverse quark motion gives   possible ways
of regularizing these end-point singularities. In this report, the
MPA is  used to study amplitudes of vector meson electroproduction
for longitudinally and transversely polarized photons within the
2G and GPD models. Singularities in  the amplitudes occurring in
collinear approximation are regularized by the transverse quark
momentum.

The leading twist term of the wave function gives a vanishing
contribution to the amplitudes with a transversally polarized
vector meson in the massless limit. To calculate these amplitudes,
it is necessary to include in consideration the higher twist terms
in the wave function. In this report, we use the $k$- dependent
wave function \cite{koerner}
\begin{equation}\label{psi}
\hat \Psi_V= g[ (\,/\hspace{-2.2mm} V+M_V) \,/\hspace{-2.9mm}E_V
                 + \frac{2}{M_V}\, /\hspace{-2.2mm} V \,
                 /\hspace{-2.9mm}
                 E_V /\hspace{-2.9mm} K
         - \frac{2}{M_V} (\, /\hspace{-2.2mm} V - M_V) (E_V \cdot K)]
         \phi_V(k_\perp^2,\tau).
\end{equation}
Here $V$ is a  momentum and $M_V$ is a mass of a vector meson,
$E_V$ is its polarization, $\tau$ is a fraction of momentum $V$
carried by the quark, and $K$ is its transverse momentum:
$K^2=-k_\perp^2$. The first term in (\ref{psi}) represents the
standard wave function of the vector meson. The leading twist
contribution to the longitudinal vector meson polarization is
determined by the $M_V \,/\hspace{-2.9mm}E_V$ term in (\ref{psi}).
The $k$- dependent terms of the wave function are essential for
the amplitude with transversely polarized light mesons.

Let us consider  vector meson production in MPA within the 2G
model. The leading over $s$ term  of the $\gamma^\star \to V$
amplitude is mainly imaginary.  The imaginary part of the
amplitude can be written as an integral over $\tau$ and $k_\perp$
and has the form \cite{martin,golos03}
\begin{equation}\label{Agv}
T_{\lambda_V,\lambda_\gamma}^V = N\,\int d\tau \int d k_\perp^2
 \frac {H^g(\xi,\xi,t)\, \phi_V(k_\perp^2,\tau)\,
A_{\lambda_V,\lambda_\gamma}(\tau,k_\perp^2)}
 {(k_\perp^2+\bar Q^2)^3},
\end{equation}
where    N is the normalization constant, $\bar Q^2=\tau \bar \tau
Q^2$, $\bar \tau =1-\tau $. Positive proton helicities are omitted
here for simplicity.  In calculation of (\ref{Agv}) the Feinman
gauge is used and $t$- channel gluons are polarized
longitudinally. The function $H^g(\xi,\xi,t)$ is connected to the
gluon GPD at $x= \xi$ point \cite{golos03}, where skewness $\xi$
is related to Bjorken-$x$ by $\xi\simeq x/2$. The meson wave
function $\phi_V$  is used in a simple Gaussian form \cite{jak93}
 \be
          \phi_V(k_\perp^2,\tau)\,=\, 8\pi^2\sqrt{2N_c}\,  a^2_V
       \, \exp{\left[-a^2_V\, \frac{\vk^{\,2}}{\tau\bar{\tau}}\right]}\,.
\label{wf} \ee Transverse momentum integration of (\ref{wf}) leads
to the asymptotic form of a meson distribution amplitude
$\phi_V^{AS}=6 \tau \bar{\tau}$.

 The hard amplitudes $A_{\lambda_V,\lambda_\gamma}$ in (\ref{Agv})
are calculated perturbatively. The  $\gamma _{L}^{\ast
}\rightarrow V_{L}$ amplitude  has the form \cite{golos03}
\begin{equation}\label{al}
A_{L,L}=4 \frac{s}{\sqrt{Q^2}} \left[\bar Q^2 + k_\perp^2 (1-4
\tau \bar \tau )\right] \,\left(\bar Q^2 + k_\perp^2 \right).
\end{equation}

For the amplitude with transversely polarized photons and vector
mesons we find
\begin{equation}\label{att}
A_{T,T}\sim \frac{2 s \, }{M_V}
 \bar Q^2\left[k_\perp^2 (1+4 \tau \bar \tau) +2 M_V^2 \tau
\bar \tau \right]\, (E^\gamma_\perp E^V_\perp).
\end{equation}
For the light meson production the resulting amplitude  is
proportional to $k_\perp^2$. The term proportional to $M_V^2$
appears in the amplitude for heavy mesons too.

The  $\gamma _{T}^{\ast }\rightarrow V_{L}$ transition amplitude
is determined by the function
\begin{equation}\label{atl}
A_{L,T}\sim \frac{2 s \, }{M_V}
 \bar Q^2\left[2 M_V^2 \tau \bar \tau-k_\perp^2 (1-2 \tau ) \right]\,
\frac{(E^\gamma_\perp r_\perp)}{M_V}.
\end{equation}
It can be found that if we omit the $k^2$ terms in the denominator
of (\ref{Agv}), the $T_{T,T}$ and $T_{L,T}$ amplitudes will have
the end-point singularities at $\tau (\bar \tau) =0$
\cite{golos03}. All amplitudes in the 2G model are mainly
imaginary. The real part of the amplitude can be obtained from the
imaginary part using the derivative rule
\begin{equation}\label{re}
{\rm Re}\; T \sim - \frac{\pi}{2}\, \frac{d}{d \ln{x}}\,
            {\rm Im}\; T\,.
\end{equation}
The real parts of the amplitudes are small, about $30\%$ with
respect to its imaginary part.

The vector meson electroproduction can be studied within the GPD
approach at large photon virtuality $Q^2$. At small Bjorken-$x$ we
shall consider as before the predominated gluon contribution. The
$\gamma _{L}^{\ast }\rightarrow V_{L}$, $\gamma _{T}^{\ast
}\rightarrow V_{T}$, $\gamma _{T}^{\ast }\rightarrow V_{L}$
amplitudes are calculated  within the MPA. In the GPD model we
consider the Sudakov suppression of large quark-antiquark
separations. These effects provide additional suppression of
contributions from the end-point regions, in which one of the
quarks entering into the meson wave function becomes soft and
 factorization breaks down. As previously, including the
transverse quark momenta regularizes singularities and gives a
possibility of calculating  the transition amplitudes at large
$Q^2$ which are important for polarized observables. The
amplitudes {$\gamma^*_\mu \,p \to V_{\mu'} p$}  ~ can be
represented in the form \cite{golsp04}:
\begin{eqnarray}\label{amptt-nf-ji}
  T_{\mu',\mu} &=& \frac{e}{2}\, C_V
         \int_0^1 \frac{d \bar x}
        {(\xb+\xi) (\xb-\xi + i\hat{\veps})}\nn\\
        &\times& \left\{\, \left[\,{\cal H}^{(g)}_{\mu'+,\mu +}\,
         +  (-1)^{\mu'+\mu}\,{\cal H}^{(g)}_{-\mu'+,-\mu +}\,\right]\,
                                   H^g(\xb,\xi,t)  \right\}\,,
\end{eqnarray}
The flavor factor for $\rho$ -meson production is
$C_{\rho}=1/\sqrt{2}$.

The hard scattering amplitudes ${\cal H}$  in (\ref{amptt-nf-ji})
are written for the positive transverse gluon polarization and can
be represented as a convolution of the hard part
$A_{\mu',\mu}^{(g)}$, which is calculated perturbatively, and the
wave function (\ref{wf}) \be {\cal H}^{V(g)}_{\mu'+,\mu +}\, =
\,\frac{2\pi \als(\mu_R)
           f_V}{N_c} \,\int_0^1 d\tau\,\int \frac{d^2\vk}{16\pi^3}
            \phi_V(k_\perp^2,\tau)\; A_{\mu',\mu}^{(g)}(x,\xi,\vk,Q^2)\,.
\ee Here  the scale $\mu_R$ is determined by the largest mass
scale appearing in the hard scattering amplitude: $\mu_R=
\mbox{max}\{\tau Q, \, \bar \tau Q,\, ...\}$.

\section{Amplitude structure and description of experiment}
The GPD model leads to the following form of helicity amplitudes
\begin{equation}
T_{LL} \propto 1\, ;\;\;\; T_{TT}^{V(g)} \propto
\frac{|\vk|}{Q}\,;\;\;\; T_{TL}^{V(g)} \propto
\frac{\sqrt{-t}}{Q}\,.
\end{equation}
This behavior is similar to those obtained in the 2G model.

The 2G and GPD approaches are  hoped to be equivalent  at small
$x$. Unfortunately, the amplitude structure in the models are not
equivalent. As  mentioned before, in the 2G model all amplitudes
are mainly imaginary. In the GPD approach  the integration over
$x$ occurs in (\ref{amptt-nf-ji})

\begin{eqnarray}\label{intx}
T^{V(g)} &\sim& \int_0^1 \frac{d \bar x\, H(\bar x)}
        {(\xb+\xi) (\xb-\xi + i\hat{\veps})}= I(\bar
x<\xi)+I(\bar x>\xi)\nonumber\\  &=&  \int_0^\xi \frac{d \bar
x\,H(\bar x)}
        {(\xb+\xi) (\xb-\xi + i\hat{\veps})}+\int_\xi^1 \frac{d \bar x\,H(\bar x)}
        {(\xb+\xi) (\xb-\xi + i\hat{\veps})}
\end{eqnarray}
For the  nonflip $T_{LL}$ and $T_{TT}$  amplitudes  we have no
singularities in integrated functions $H(\bar x)$, and both
$I(\bar x<\xi)$ and $I(\bar x>\xi)$ contribute to the Re part of
amplitude. These integrals are not small, have different signs and
compensate each other mainly. As a result, the real part of the
$LL$ and $TT$ amplitudes is quite small and is consistent with the
one obtained from (\ref{re}). In the case of the $T_{LT}$
amplitude we have quite a different result. In this case,  we find
an additional coefficient $1/\sqrt{s\, u} \propto
1/\sqrt{\xb^2-\xi^2}$ in the hard amplitude $H$ in (\ref{intx})
which becomes imaginary in the $\xb<\xi$ integration region.
Consequently,  the real part of the $T_{LT}$ amplitude is
determined only by $I(\bar x>\xi)$ integral. This contribution is
not small and we find that $\mbox{Re }T_{LT}> \mbox{Im }T_{LT}$
for this amplitude. Thus, properties of the $T_{LT}$ amplitude in
the 2G and GPD models are quite different. It is difficult to
imagine that these amplitudes might be equivalent at small $x$.
\vspace{-3mm}
\begin{figure}[h]
\begin{center}
\includegraphics[width=.44\textwidth,bb=61 358 552 739]{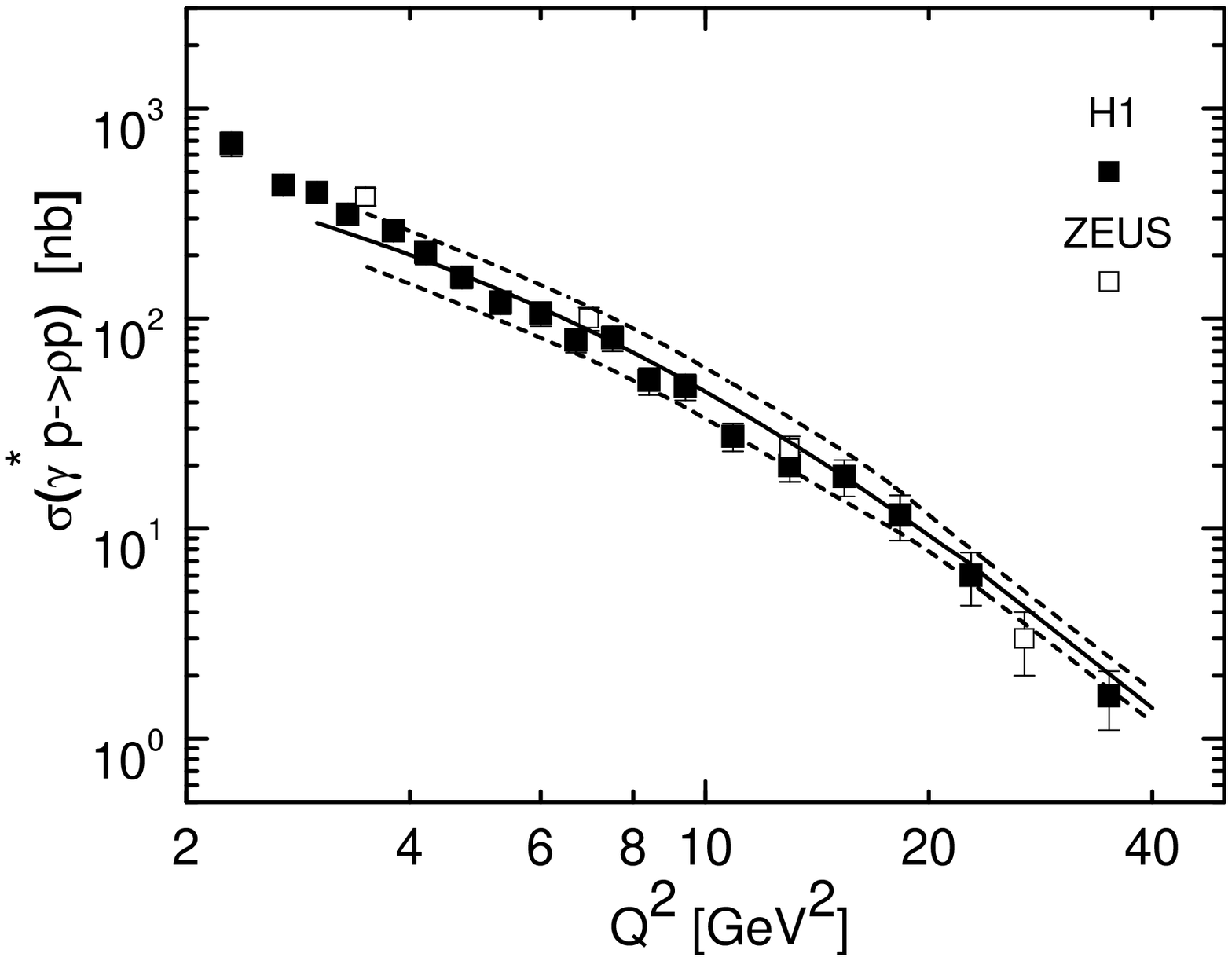}
\includegraphics[width=0.44\textwidth, bb= 35 358 500 725]{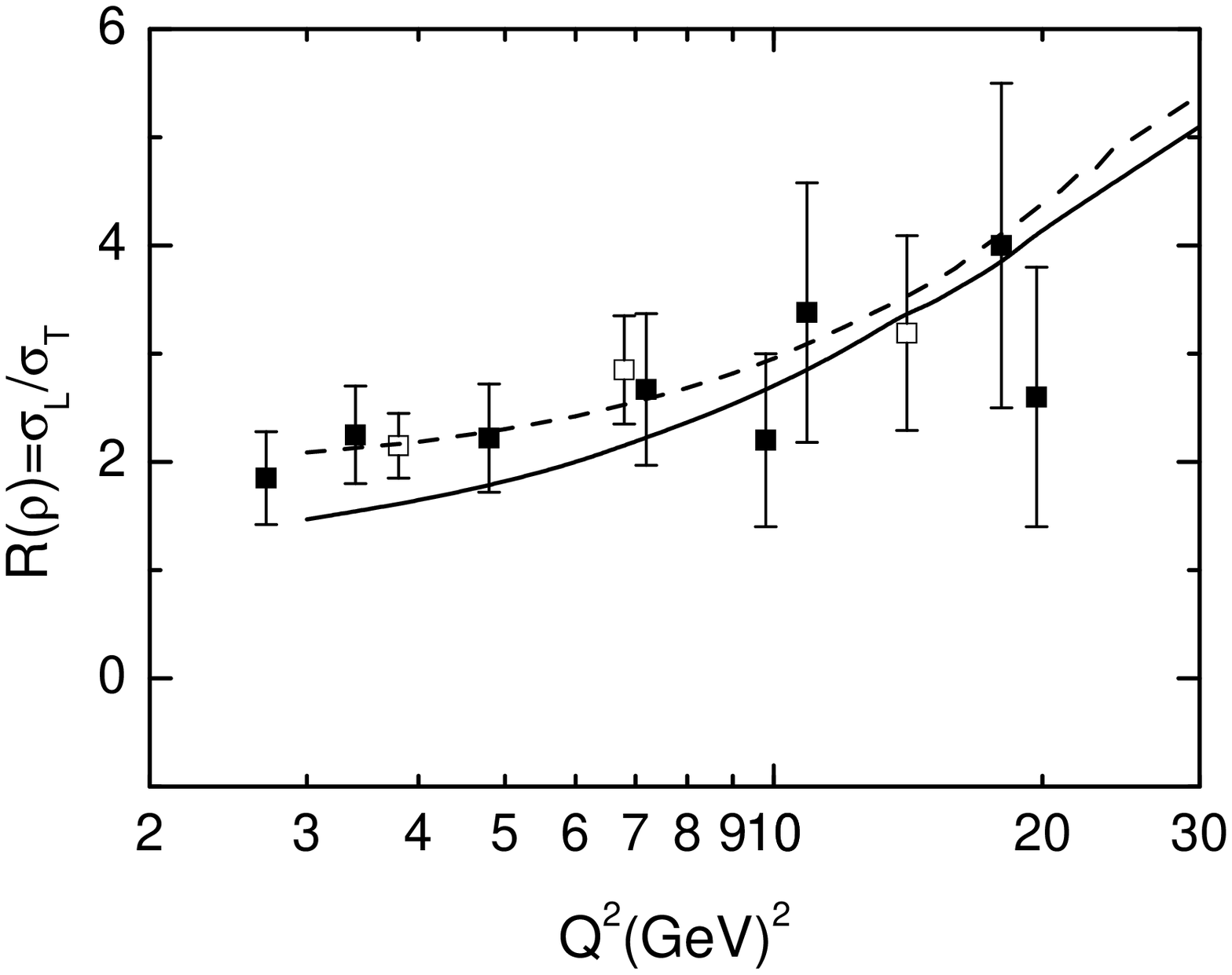}
\end{center}
\caption{Left: The cross section for $\gamma^*\, p\to \rho^0\, p$
vs. $Q^2$  for fixed values of $\langle W\rangle=75\gev$. Full
line- GPG model results. Dashed lines show the $\mu_R$
sensitivity. Data are  from \cite{h1,zeus}.} \caption{Right: $Q^2$
dependence R of $\rho$ production at $\langle W\rangle=75\gev$.
Full curve -2G model, dashed curve -GPD results. Data are from
\cite{h1,zeus}.}
\end{figure}

Let us consider the description of experimental data  in the 2G
and GPD models. In both the cases we have the $a_v$ parameter in
the wave function which determines the average value of $\langle
k_\perp^2\rangle$ in hard subprocess.  In the numerical evaluation
of meson electroproduction a reasonable description of
experimental data is obtained for $ a_\rho= 0.8\, \gev^{-1}$ in
the 2G model and for $a_\rho= 0.52\, \gev^{-1}$ in the GPD model.
The parameter $f_v$ is determined by the standard value and for
$\rho$ meson production we use $f_\rho = 0.216\,\gev$. Estimations
of the amplitudes are carried out using the $\Lambda_{QCD}=0.22
\gev$. The cross section for $\gamma^* p \to \rho p$
 production integrated over $t$  is shown
in Fig. 1 (full line). Good agreement with experiment is to be
observed. The results for $\phi$ production can be found in
\cite{golsp04}. It is important to analyse the dependence of cross
section on the scale $\mu_R$. The results for cross section for
$\tilde \mu_R= \{\sqrt{2}\mu_R,\mu_R/\sqrt{2}\}$ are shown in
Fig.1 by dashed lines. It can be seen that the $\tilde \mu_R$
sensitivity of the cross section is of the order of experimental
errors.

Using the calculated amplitudes we can determine contributions to
the cross section with longitudinal and transverse photon
polarization and  its ratio as
\begin{equation}\label{nlt}
N_L=  |T_{LL}^{V}|^2\,,\;\;\;\; N_T=  |T_{TT}^{V}|^2+
|T_{LT}^{V}|^2\,,\;\;\;\; R=\frac{N_L}{N_T}.
\end{equation}
Note that in (\ref{nlt}) summation over proton helicities is
assumed. We omit here the $T_{TL}$ and $T_{-TT}$ amplitudes which
are small in the models. In terms these quantities the
spin-density matrix elements (SDME) can be defined, e.g.
\begin{equation}\label{sdme}
r_{00}^{04} = \frac1{N_T+ \varepsilon N_L}\,
      (|{T_{LT}^{V}|^2\,+\varepsilon\,|T}_{LL}^{V}|^2).
\end{equation}
\begin{figure}[h]
 \epsfysize=78mm
 \centerline{\epsfbox{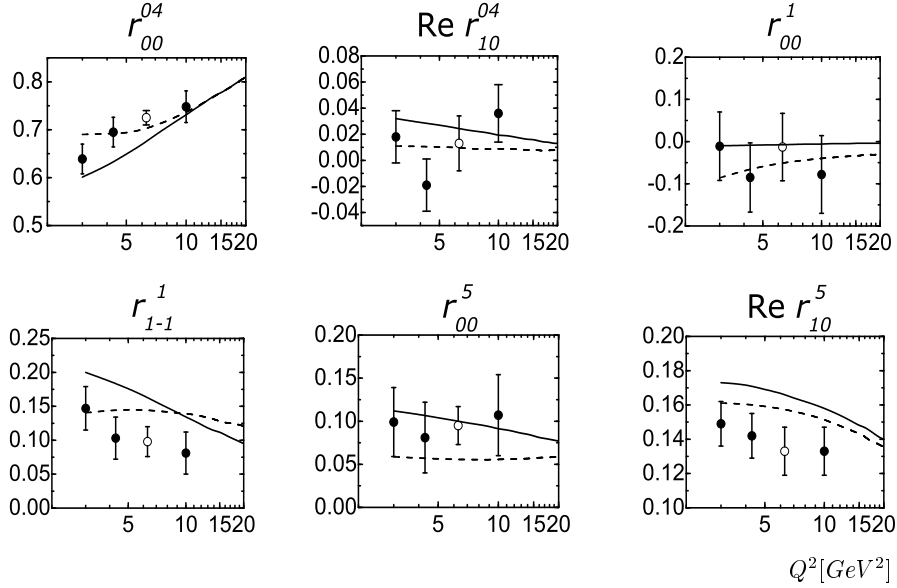}}
 \caption{$Q^2$ dependence SDME of $\rho$ production at $\langle -t\rangle=0.15\,\gev^2$
 and $\langle W\rangle=75\gev$. Full curve -2G model, dashed curve -GPD results.
 Data are taken from \cite{h1,zeus}.}
\end{figure}
The model results for the ratio of cross section $R$ are shown in
Fig. 2. For both the models this ratio is growing with $Q^2$ and
in consistent with experiment. In Fig.3, we show six essential
SDME. In the approximation, when we put  the $T_{TL}$ and
$T_{-TT}$ amplitudes to be zero, the other SDME are connected with
the matrix elements from  Fig. 3 or equal to zero. The description
of experimental data in  both the models is reasonable.

We would like to note that from the data on $d\sigma/dt$ the
diffraction peak slope $B \sim 6\gev^{-2}$ at $Q^2 \sim 5\gev^2$
can be determined \cite{h1,zeus}. This value is connected with the
diffraction peak slope of the $T_{LL}$ amplitude because its
contribution to the cross section is most essential. The
diffraction peak slopes of the $T_{TL}$ and $T_{TT}$ amplitudes
are not well defined. In calculation of spin observables we
suppose that the diffraction peak slope $B_{LT} \sim B_{LL}$ and
 $B_{TT}$ might be different. The slope $B_{TT} \sim B_{LL}$ in
the 2G model and $B_{TT} \sim B_{LL}/3$ in the GPD model is used.
Predictions of both the models are in agreement with the known $t$
-dependence of experimental data at small momentum transfer
\cite{h1}. The results found in \cite{2gm} are very close to
estimations obtained here within the 2G model (Fig.3).

\section{Conclusion}
Light vector meson electroproduction at small $x$ was analyzed in
this report within the 2G and GPD models. In both the models the
amplitudes were calculated  using MPA and the wave function
(\ref{psi}) which consider the transverse quark momentum. By
including  the higher twist effects $k_\perp^2/Q^2$ in the
denominators of $T_{\lambda_V,\lambda_\gamma}^V$ in (\ref{Agv}) we
regularize the end-point singularities in the amplitudes with
transversally polarized photons. It was found that the 2G and GPD
models, which are expected to be equivalent at small $x$, lead to
similar results for the leading twist $T_{LL}$ amplitude. At the
same time, properties of the amplitudes suppressed as a power of
$1/Q$ are different  in the models. This was demonstrated here for
the $T_{LT}$ amplitude. Thus, the 2G and GPD models are not
completely equivalent at small $x$. It was shown that the
diffraction peak slopes of the $T_{TT}$ and $T_{LT}$ amplitudes
are not well defined. The knowledge of these slopes is essential
in analyses of SDME. Information about $B_{TT}$ and $B_{LT}$ can
be obtained from $t$ -dependence of SDME.

 At the same time, both approaches lead to an accurate
description of the cross section for the light meson production.
We found a reasonable results for SDME and $R$ ratio in the 2G and
GPD models. This means that at the present time we have two
solutions for the scattering amplitudes which are in agreement
with existing experimental data. Unfortunately, all data on spin
observables have now large experimental errors. This does not
permit one to determine which model is relevant to experiment. To
clarify the situation, an additional theoretical  study of the
$T_{TT}$ and  $T_{LT}$ amplitudes is needed. An experimental
investigation to reduce errors in SDME is extremely important.
Study of $t$ dependence of SDME can give  important information on
either diffraction peak slopes in helicity amplitudes are of the
same order of magnitude or different.\\

This work is supported  in part by the Russian Foundation for
Basic Research, Grant 03-02-16816  and by the Heisenberg-Landau
program.

\end{document}